\documentstyle[twocolumn,epsf,aps]{revtex}
\draft
\begin{document}
\title{Self-organizing domain structure in a driven lattice gas}
\author{Gy\"orgy Szab\'o, Attila Szolnoki, Tibor Antal, and Istv\'an
Borsos}
\address
{Research Institute for Materials Science, H-1525 Budapest,
POB 49, Hungary}
\address{\em \date{\today}}
\address{
\centering{
\medskip \em
%\begin{abstract}
\begin{minipage}{15.4cm}
{}~~~Instead of the homogeneous ordered particle distributions
characteristic to equilibrium systems a self-organizing polydomain
structure is found to be stable at low temperatures in a square
lattice-gas model with repulsive nearest neighbor interaction when
the particle jumps are biased by a uniform electric field. We have
performed Monte Carlo simulations to study the domain structure
varying the temperatures, fields and system sizes.  Revision of the
traditional picture on this system raises interesting problems as
discussed.
%\end{abstract}
%
\end{minipage}
}}
\maketitle
\narrowtext

The introduction of driven lattice gases \cite{kls} was motivated
by the demands to study non-equilibrium (open) systems which
exhibit phase transitions and seem to be tractable with the improvement
of dynamical methods using the concepts of equilibrium statistical
physics. In these lattice-gas models the thermal jumps of interacting
particles are biased by a uniform electric field $E$ resulting in a
stationary particle transport through the system if periodic boundary
conditions are used. These systems exhibit many interesting
features (for a review see \cite{schm,konyv}). Obviously,
the equilibrium behavior should be reproduced in the
limit $E \to 0$. From practical point of view these models are
able to describe the effect of driving field on the ordering
processes in superionic conductors \cite{sic1,sic2}.

Now our consideration will be restricted to a half-filled driven lattice
gas on a square lattice with repulsive nearest neighbor interaction.
Accepting the previous notations the nearest neighbor interaction
is considered as an energy unit in which both the temperature
($k_B=1$) and strength of driving field are measured
\cite{kls,schm}. The system has a twofold degenerate ground
state in which the particles form chessboard-like ordered structure.
In equilibrium one of this structure will appear as a result of a
sublattice ordering when cooling the system through the N\'eel
temperature $T_N$. The feature of the corresponding critical
transition is well described in the literature \cite{stan}. The early
investigations apparently confirmed the naive expectation, namely,
the universal behavior of the ordering process remains unchanged
in the presence of a weak driving field. More precisely, dynamical
mean-field analysis \cite{dickr}, field theoretical investigation
and Monte Carlo (MC) simulations \cite{lsz} have suggested that
the N\'eel temperature decreases with $E$ and the continuous
transition becomes first order above a threshold value. These
analyses have shown that the order parameter as a function of
temperature depends on the jump rate (Metropolis or Kawasaki)
if $E>0$. In both cases the transition temperature vanishes
for $E=2$.

In a previous paper \cite{polyd} the MC simulations were
reinvestigated using as large as $300 \times 1500$ system size
for a fixed driving field ($E=0.4$) where continuous transition
was predicted. These simulations have demonstrated clearly
that a self-organizing polydomain structure characterizes the
stationary state at low temperatures.
In other words, the applied field prevents both
the spontaneous symmetry breaking and the related critical
phenomena.

In this Letter we look over the recent results and give further
evidences to support the former picture. Our main goal is to
illuminate the variety of phenomena and emphasize the joints
with other areas. First we briefly review the results of a
phenomenological model which clarifies the appearance of some
typical lengths leading to a size effect. It will be shown that this
size effect resolves the discrepancy between the
MC data obtained for small and large system sizes. We could
distinguish three different types of domain patterns when varying
the temperature, field strength and system size. The domain size
distribution shows power law behavior in the self-organizing
polydomain structure considered as ``thermodynamic limit''.
Finally we summarize our main conclusions.

The visualization of particle distribution during the MC simulations
at low temperatures makes clear that most of the jumps take
place at the interfaces separating the ordered regions.
Consequently, the particle transport is also
localized along the interfaces in the presence of a driving field.
For curved interfaces the induced current results in an accumulation
of extra particles (holes) at the parts where the curvature is negative
(positive).
The interface is driven by the field with a velocity propoportional
to the charge density and perpendicular field component.  In the
mentioned phenomenological model the interface shape, the particle
density and current along the interface are described by single valued
functions of time and $x$ coordinate \cite{polyd}. Beside the
mentioned phenomena the deterministic equations of motion take
the effect of surface tension into consideration as suggested for
equilibrium systems \cite{ac}. In this model the time evolutions
of the interface and charge density are separable if $E=0$.
Namely, the charge fluctuations along the interface
are dumped by a diffusion process while the interface motion is
controlled by the surface tension.

Although in the presence of a field the situation is more complicated
because of the coupling, the equations of motion have trivial solutions
corresponding to a neutral, tilted, standing, planar interfaces.
Excepting the interfaces parallel to the field, these solutions
are found to be unstable against periodic perturbations in
agreement with MC simulations \cite{frac}. According to a
linear stability analysis the amplitude increases exponentially
with time if the wave-length exceeds a treshold value. The
amplification rate has a maximum at a given wave-length
($\lambda^{\star}$) proportional to $1/E$. Following the
initial (exponential) increase of the periodic components there
appears a finger formation controlled by non-linear effects.
In this case the typical finger width can be estimated as
$\lambda^{\star}/2$.  Consequently, the growing domains
are splitted into strips whose length depends on the thermal
fluctuations. Finally the system evolves into an anisotropic,
self-organizing domain structure at sufficiently low temperatures.
In this stationary state some parts of  the moving interfaces can
meet and annihilate each other while the accumulated charges are
neutralized (in part). This process unites two distinct domains
and generates extra defects into the bulk phase. All these
phenomena can be observed clearly when displaying the
particle distribution during the MC simulations.

The above interfacial instability is analogous to the
Mullins-Sekerka instability observed in crystallization
\cite{ms,langer}. Furthermore, a similar phenomenon is found
in the driven lattice gas with attractive interaction for the
interfaces separating the high- and low-density phases
\cite{leung,yeung,szabo}. It is emphasized that the interfaces
parallel to the field remain stable in the driven diffusive
systems for both the attractive and the repulsive interactions.

Beside the preservation of this self-organizing stationary state
the enhanced interfacial particle transport breaks up the
monodomain structure into a polydomain one via a nucleation
mechanism as observed in MC simulations \cite{polyd,teitel}.
Due to thermal fluctuations ``islands'' of the opposite phase are
formed in the homogeneous initial state. These ``islands''
polarized by the above mechanism  elongate along the field.
Using the mentioned phenomenological model we could
determine the variation of the domain area. For simplicity
we have studied a circular domain and the charge distribution
along its boundary is chosen to be a stationary solution for fixed
radius. It is found that the area increases if the radius exceeds a
critical value, $R_c$ which is proportional to $1/E$.

In accordance with the above picture we can distinguish three
different charcteristic lengths related to the interfacial phenomena;
i.e., the ``typical width and length of strips'' and the critical
nucleon size. The MC simulations show that $R_c$ and
$\lambda^{\star}/2$ are comparable while the ``typical strip
length'' (or longitudinal correlation length) may become
significantly larger when decreasing the temperature for a
fixed field ($E=0.4$ in Ref.~\cite{polyd}).

The existence of these characteristic lengths leads to an obvious
size effect. The self-organizing domain structure is preserved by
the interfacial instability when the system sizes are chosen to be
much larger than these former values.  On the other hand, if the
MC simulation is performed on a small system then one can
observe an ordered  monodomain structure at low temperatures
and phase transition can be concluded. For example, the finite size
scaling of the order parameter can be performed for $L\le 40$ and one can
deduce a ``critical temperature'' $T_N(E=0.4)=0.488$. For larger $L$,
however, the scaled data deviate significantly from the scaling function.

\begin{figure}
\centerline{\epsfxsize=8cm
                   \epsfbox{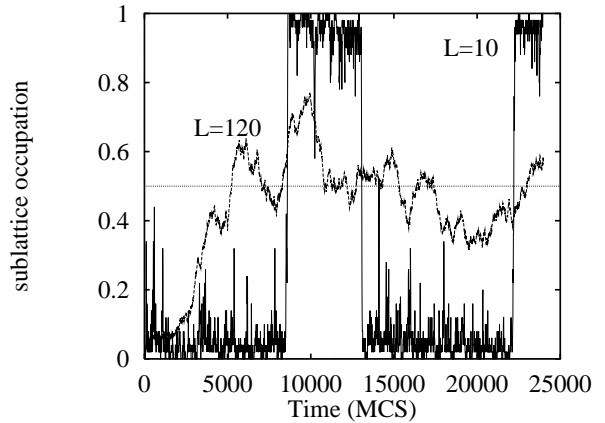}
                   \vspace*{2mm}         }
\caption{Time dependence of sublattice occupation
in MC simulations performed at $E=0.4$ and $T=0.48$ for
different $L$.}
\label{fig:nt}
\end{figure}

In small system the ordered stucture evolves into its
counterpart via the nucleation mechanism as demonstrated in
Fig.~\ref{fig:nt} (for $L=10$) where the sublattice occupation
is plotted as a function of time measured in Monte Carlo steps
per particles (MCS).

{}From Figure \ref{fig:nt} it is clear that the evolution of
sublattice occupation shows different behavior for $L=120$ if
the system is started from an ordered structure. Evidently, the
fluctuation of sublattice occupation (around $1/2$) decreases
when choosing even larger systems.

Beside the mono- and polydomain structures we can distinguish
a third type of domain patterns when the system size equals
approximately with the longitudinal correlation length which
is generally larger then $\lambda^{\star}/2$ (or $R_c$). In this
case closed strips are formed with interfaces parallel to the field.
These interfaces are not affected by the instability mentioned
above. Due to the thermal fluctuations the interfaces move
randomly, the neighboring ones can meet and annihilate each
other. The ``multistrip'' state can be considered as a stationary
one because the nucleation mechanism retrieves the loss caused
by the annihilation.

\begin{figure}
\centerline{\epsfxsize=8cm
                   \epsfbox{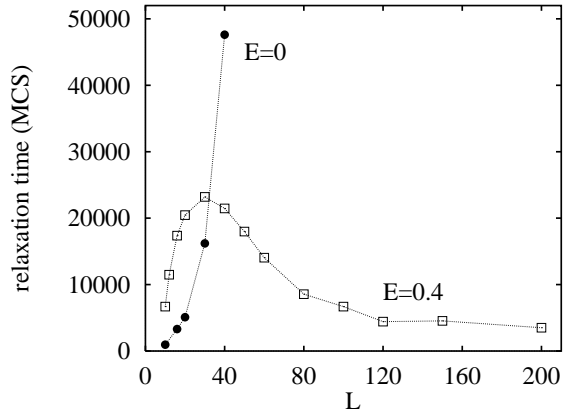}
                   \vspace*{2mm}         }
\caption{Size dependence of relaxation time in a driven system
(open squares) and for $E=0$ (closed circles) at $T=0.48$.}
\label{fig:tau}
\end{figure}

The common feature of the above situations is that the time
average of the order parameter disappears. We have introduced
a relaxation time to characterize the disappearance of this
quantity. A series of MC simulations was
performed varying the system size for fixed field $E=0.4$
and temperature $T=0.48$. The time variation of the order
parameter has been determined by averaging over 150 MC runs
when the system is started from one of the ordered structure.
Following the method introduced  by Binder and
M\"uller-Krumbhaar \cite{bmk} we have evaluated the
relaxation time for different system sizes
(see Fig.~\ref{fig:tau}).

For small sizes the relaxation time increases with $L$. The curve
has a maximum for $L \approx 36$ and tends to a constant value
when $L \to \infty$. As a comparison with the equilibrium
system we have repeated these calculations for $E=0$ and
$L \le 40$. In contrary to the driven system the relaxation
(ergodic) time increases monotonously with $L$ in the
absence of drive as demonstrated. Unfortunately, the systematic
analysis becomes time consuming for lower temperatures
because of the rapid increase of relaxation time.

\begin{figure}
\centerline{\epsfxsize=8cm
                   \epsfbox{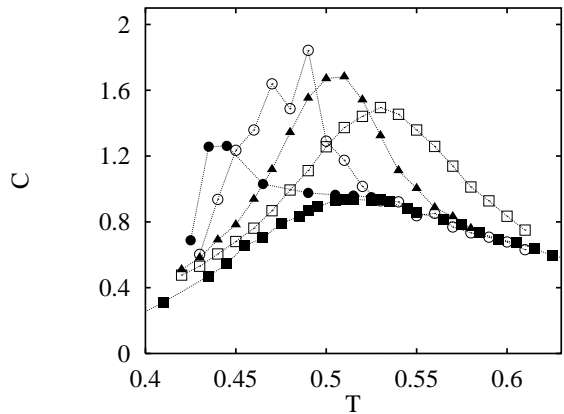}
                   \vspace*{2mm}         }
\caption{Specific heat vs. temperature at $E=0.4$ for different
system sizes $L=16$ (open squares), 32 (closed triangle), 64
(open circle), 128 (closed circles) and 256 (closed squares).}
\label{fig:sh}
\end{figure}

The interfacial energy gives significant contribution to the total
energy in the driven sytem. Consequently the size effect modifies
the specific heat drastically as shown in Fig.~\ref{fig:sh}.
The peak increases and moves to left when choosing $L=16$,
32 and 64. For larger systems, however, the appearance of
interfaces decreases the peak height. Notice that
the simulations obtained for $L=256$ reproduce the previous
data \cite{polyd}.

\begin{figure}
\centerline{\epsfxsize=8cm
                   \epsfbox{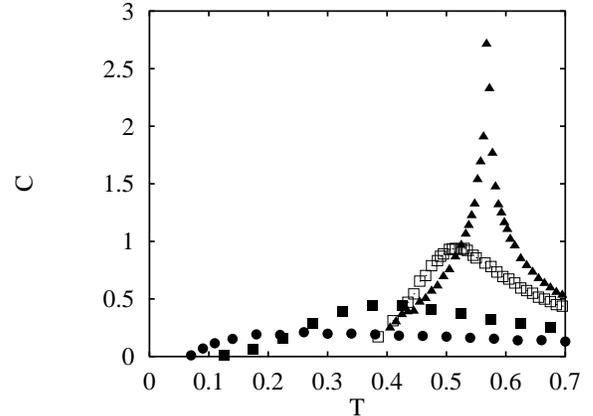}
                   \vspace*{2mm}         }
\caption{Specific heat for different fields: $E=0$ (triangles),
0.4 (open squares), 1 (closed squares), and 1.5 (bullets).}
\label{fig:ce}
\end{figure}

For stronger field the characteristic lengths become shorter and
the polydomain structure is more tractable. As a consequence
the MC simulations can be performed on smaller systems to
have reliable results. The simulations have justified the absence
of long range order at low temperatures. Furthermore, the
$\lambda$ singularity of specific heat spreads away as
illustrated in Fig.~\ref{fig:ce}.

These data suggest that the critical behavior appears only in
the limit $E \to 0$.
Unfortunately, the MC simulations are not adequate to
study rigorously the crossover from the smooth transition
to the critical one when decreasing $E$ because of the extremely
large system sizes required to avoid the mentioned size effects.
At the same time, the results of simulations agree quantitatively
with the predictions of a generelized mean-field analysis performed
at the level of 6-point approximation \cite{gmf} for $E>1$.

The appearance of pattern formation in the present  model is not
surprising because similar phenomena are found in some other
open systems. The best known example is the Rayleigh-B\'enard
instability \cite{RB}. The curiosity of this driven lattice gas is
related to the fact that here the pattern formation is preserved by
the enhanced interfacial particle transport. That is, an
interfacial phenomenon is capable to prevent the ordering process
implying drastical modifications in the macroscopic features.

\begin{figure}
\centerline{\epsfxsize=8cm
                   \epsfbox{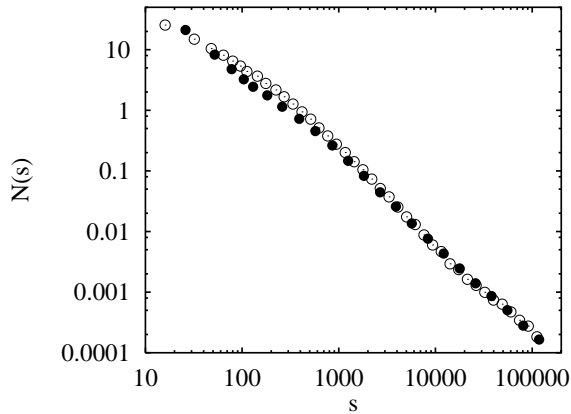}
                   \vspace*{2mm}         }
\caption{Log-log plot of the average number of domains vs. their
size for a self-organizing polydomain structure (closed circle) and
for a state (open circle) during the roughening process.}
\label{fig:dsd}
\end{figure}

The polydomain structure consisting of two phases is
topologically equivalent to a pattern like ``droplets inside of
droplets inside droplets ...'' characteristic of the critical domain
picture of Ising model \cite{droplet}. According to the
deterministic interfacial evolution model the appearence of
characteristic lengths leads to minimum size for droplets.
In the real system the smaller droplets appear as a
consequence of thermal fluctuations. The resultant domain
structure is topologically similar to a quenched state
roughened for a given time. To demonstrate
it we have determined the domain size distribution functions
in both cases. Closed circles in Fig.~\ref{fig:dsd}
show the average number of domains of size $s$ in a rectangular
system with $500 \times 1500$ sites for $E=0.4$ and $T=0.4$.
Here the open circles are obtained by averaging
over 1000 patterns developed from a random distribution after
a thermalization of 300 MCS in the absence of driving field
for a fixed temperature $T=0.4$. In this former case there
also exists a characteristic domain size increasing
monotonously with time.

The domain size distribution functions are generally sensitive
to the choice of criteria defining domains and interfaces.
In the above comparison we used the same criteria to
demonstrate the similarity between the two types of states.
Preliminary results suggest that this behavior is not affected
by the anisotropy. Further systematic analysis
is required to have quantitative conclusions.

For weak fields the interfacial phenomena including noises
affect dominantly the long range order while it allows
nearly perfect order inside the domains.
In this limit the polydomain state can be considered as a
noisy surface evolution problem. Such an approach raises many
questions not yet investigated in the area of surface growth
\cite{barab}. For example, this technique may give
some predictions for the longitudinal domain (or correlation)
length.  Furthermore, it would be interesting to classify
the interface evolution models which are able to preserve
such a self-organizing domain structure.

In summary, at low temperatures the present driven
lattice-gas model exhibits a self-organizing polydomain
structure which is topologically similar to those appearing
during the roughening process. The interfacial effects not
only preserve the self-organizing process but they are able
to destroy the monodomain structures. The related characteristic
lengths are responsible for the significantly different features
found for the small and sufficiently large systems. The
temperature- and field-dependence of the self-organizing
state raises many questions to be studied in the near future.

\acknowledgements

This research was supported by the Hungarian National Research
Fund (OTKA) under Grant Nos. T-4012 and T-16734.

\end{document}